\documentclass[11pt,twoside]{article}
\usepackage{asp2004}
\usepackage{epsf}
\usepackage{epsfig}
\usepackage{lscape}
\usepackage{graphicx}
\usepackage{longtable}
\begin{document}
\title{Standard stars for linear polarization observed with FORS1}
\author{
L.~Fossati$^{1,2}$,
S.~Bagnulo$^{2}$,
E.~Mason$^{2}$,
E.~Landi Degl'Innocenti$^{3}$}
\affil{
$^{1}$Institut fuer Astronomie, Universitaet Wien, Tuerkenschanzstrasse 17,
      A-1180 Wien, Austria\\
$^{2}$European Southern Observatory, Alonso de Cordova 3107, Vitacura, Santiago, Chile\\
$^{3}$Dipartimento di Astronomia e Scienza dello Spazio, Universit\`a di Firenze,
      Largo Enrico Fermi 2, I-50125 Firenze, Italy
}
\begin{abstract}
We present the analysis of the observations of standard stars for
linear polarization obtained from 1999 to 2005 within the context of
the calibration plan of the FORS1 instrument of the ESO VLT. We have
considered observations carried out both in imaging polarimetric and
in spectropolarimetric mode. Broadband polarization was obtained in
the Bessel \textit{BVRI} filters; spectropolarimetry was obtained with
various grisms covering different optical ranges and with a typical
resolution of a few hundreds. Spectropolarimetric data have been
convolved with the transmission functions of the Bessel filters, which
enabled us to calculate polarization values equivalent to broadband
polarization measurements in imaging mode. Finally, for each star,
instrument mode, and for each Bessel filter band, we have calculated an
average polarization value obtained after filtering all available data
with a $k\sigma$-clipping algorithm.
\end{abstract}
\section{Introduction}
When taking polarimetric measurements of scientific targets,
observations of standard stars for linear polarization are of crucial
importance to calibrate and monitor instrument performances. Popular
lists of standard stars are, e.g., those by Serkowski
\cite{Ser74} and Hsu and Breger \cite{HsuBre82}. Unfortunately, most of
the `classical' standard stars for polarization presented in these
works are too bright for the instruments of the large size class
telescopes.

At the ESO Very Large Telescope (VLT), observations of standard stars
deemed to exhibit either zero polarization or substantial polarization are
routinely performed within the context of the FORS1 calibration
plan. Usually, a star known to exhibit a large signal of linear
polarization is observed during those nights when a science target is
observed in polarimetric mode. Occasionally, a non polarized star is
also observed. However, the stability of these stars used as
standard for linear polarization has never been extensively checked so
far. Therefore, we have decided to retrieve from the archive a large
sample of observations of polarimetric standard stars obtained with
FORS1 and to check their consistency.
\section{Instrument and observations}
FORS1 is the visual and near ultraviolet {\bf FO}cal {\bf R}educer and low
dispersion {\bf S}pectrograph mounted at the Cassegrain focus of one of the
four 8\,m units of the ESO VLT, and works in the wavelength range
$330-1100$\,nm. FORS1 is equipped with polarization analyzing optics,
which include a half-wave and a quarter-wave
retarder plates, both superachromatic
and a Wollaston prism with a beam divergence of 22''.
This allows the measurement of linear and circular polarization, both
in direct imaging and spectroscopy. In imaging polarimetric mode
(IPOL) the field of view is $6.8' \times 6.8'$ and the magnitude limit
is $R=23$, for a 1\,\% accuracy in the polarization measure, and
with 1\,h exposure time. In spectropolarimetric mode (PMOS), the
magnitude limit is between $R=17.2$ and 19.3, for a spectral
resolution between 260 and 1700 (with a 1'' slit width, depending on
the grism inserted).  The instrument is described in Appenzeller et
al.\ (1998).

We have retrieved from the ESO data archive ({\tt
http://archive.eso.org}) all observations of polarimetric standard
stars taken with FORS1 from April 1999 to March 2005. A preliminary
inspection to the data was performed to identify and discard a few
saturated exposures and some frames with poor image quality. The list
of the useful observations is presented in Table~\ref{table star
list}. About 210 observations in IPOL mode were obtained with the
\textit{BVRI} Bessel filters (observations with the \textit{U} Bessel 
filter are not possible in polarimetric mode as the filter is situated
in the same wheel as the Wollaston prism). About 130 observations in
PMOS mode were obtained with a variety of grism and filter
combinations. Both for IPOL and PMOS modes, the linear polarization
was usually measured taking a series of four observations with
the half wave retarder waveplate at $\alpha = 0^\circ$, $22.5^\circ$, $45^\circ$, and 
$67.5^\circ$, respectively, where $\alpha$ indicates the
angle between the acceptance axis of the ordinary beam of the
Wollaston prism and the fast axis of the retarder waveplate. The
acceptance axis of the Wollaston prism was always aligned to the North
Celestial Meridian, which therefore represents the reference direction
for all linear polarization measurements of this paper.

\begin{table}
\caption[ ]{List of the observed stars. Sp.\,T.\ indicates the
spectral type of the star.  $N({\tt I})$ and $N({\tt P})$ indicate the numbers of
observations in IPOL and PMOS mode respectively. Type indicates if the
star is a candidate polarized standard star (P) or unpolarized
standard star (nP)}
\label{table star list}
\vspace{5mm}

\rotatebox{90}{
\begin{tabular}{rrllrlrrc}
\hline
\hline
\multicolumn{1}{c}{RA}&
\multicolumn{1}{c}{DEC}&
  &  &  &  &  &  &  \\
\multicolumn{2}{c}{J2000}&
\multicolumn{1}{c}{Id\,1}&
\multicolumn{1}{c}{Id\,2}&
$V$& Sp.\,T.&$N$({\tt I})&$N$({\tt P})& Type \\
\hline
05:36:23.01 &    21:11:11.4 & HD\,245310   & BD+21 901         & 8.96  & B2e & 0  & 2  & P  \\
05:41:37.85 & $-$01:54:36.5 & NGC\,2024 1  & JP11 3950         & 12.17 & B0  & 15 & 12 & P  \\
06:05:05.67 &    23:23:38.5 & HD\,251204   & BD+23 453         & 10.20 & B0  & 0  & 1  & P  \\
06:06:41.04 & $-$42:17:55.7 & HD\,42078    & CD$-$42 2343      & 6.16  & Am  & 0  & 6  & P  \\
09:06:00.01 & $-$47:18:58.2 & Ve\,6$-$23   & GSC 08169$-$00417 & 12.12 & B0V & 45 & 23 & P  \\
12:51:03.56 & $-$61:14:37.7 & HD\,111579   & CD$-$60 4390      & 9.17  & B2  & 0  & 3  & P  \\
14:28:50.87 & $-$60:32:25.1 & HD\,126593   & CD$-$59 5269      & 8.67  & B1  & 0  & 1  & P  \\
16:18:43.73 & $-$51:27:57.8 & CD\,$-$51 9993&TYC 8323$-$1798$-$1& 10.25& B   & 3  & 0  & P  \\
17:12:19.95 & $-$04:24:09.3 & HD\,155528   & BD$-$04 4244      & 9.62  & B9  & 1  & 2  & P  \\
17:43:19.59 & $-$28:40:32.8 &CD\,$-$28 13479& Hilt 652         & 10.80 & B1  & 26 & 16 & P  \\
17:43:47.02 & $-$07:04:46.6 & HD\,161056   & BD$-$07 4487      & 630   & B1  & 2  & 2  & P  \\
17:45:43.70 & $-$29:13:18.1 & HD\,316232   & CD$-$29 13940     & 10.40 & O   & 9  & 6  & P  \\
18:11:58.10 & $-$14:56:09.0 & BD\,$-$14 4922& Hilt 715         & 9.73  & O9  & 11 & 7  & P  \\
18:40:01.70 & $-$12:24:06.9 & BD\,$-$12 5133& Hilt 781         & 10.40 & B1V & 31 & 9  & P  \\
18:41:22.57 & $-$13:50:28.9 & BD\,$-$13 5073& Hilt 785         & 10.40 & B1e & 4  & 2  & P  \\[3mm]
00:02:10.75 & $-$43:09:55.6 & WD\,2359$-$434 & GJ 915          & 13.05 & DAs & 9  & 0  & nP \\
01:37:18.59 & $-$40:10:38.5 & HD\,10038    & CD$-$40 404       & 8.14  & A2m & 1  & 4  & nP \\
01:57:56.14 & $-$02:05:57.7 & HD\,12021    & BD$-$02 329       & 8.85  & A0  & 0  & 1  & nP \\
02:11:16.69 & $-$46:35:06.2 & HD\,13588    & CD$-$47 663       & 7.90  & A1m & 1  & 5  & nP \\
02:35:07.60 &    03:43:56.8 & WD\,0232+035 & Feige 24          & 12.40 & DAw & 0  & 1  & nP \\
03:10:31.02 & $-$68:36:03.4 &WD\,0310$-$688& TYC 9145$-$601$-$1& 11.10 & DA  & 6  & 0  & nP \\
07:52:25.51 & $-$23:17:46.8 & HD\,64299    & BD$-$22 2058      & 10.11 & A1V & 0  & 2  & nP \\
07:53:08.38 & $-$67:47:32.2 &WD\,0752$-$676& GJ 293            & 14.09 & DC  & 9  & 0  & nP \\
11:13:50.74 & $-$52:51:21.2 & HD\,97689    & CD$-$52 4222      & 6.82  & A0m & 0  & 3  & nP \\
16:17:55.25 & $-$15:35:52.4 &WD\,1615$-$154& G 153$-$41        & 12.40 & DA  & 6  & 6  & nP \\
16:23:33.84 & $-$39:13:46.2 & WD\,1620$-$391 & CD$-$38 10980   & 11.00 & DA  & 16 & 3  & nP \\
19:02:08.52 & $-$41:54:37.8 & HD\,176425   & CD$-$42 13839     & 6.21  & A0V & 1  & 2  & nP \\
20:10:56.85 & $-$30:13:06.6 &WD\,2007$-$303& CD$-$30 17706     & 12.18 & DA  & 0  & 5  & nP \\
20:42:34.75 & $-$20:04:35.9 &WD\,2039$-$202& HIP 102207        & 12.34 & DAw & 5  & 5  & nP \\
21:52:25.38 &    02:23:19.6 & WD\,2149+021 & HIP 107968        & 12.73 & DA  & 7  & 3  & nP \\
\hline
\end{tabular}
}
\end{table}

\section{Data reduction}
Stokes~$Q$ and $U$ parameters are defined as in Landi Degl'Innocenti
et al. \cite{Lanetal07}, with the reference axis coinciding with the
North Celestial Meridian.  In the following, we will consider the
ratios $Q/I$ and $U/I$, adopting the notation
\begin{equation}
P_Q = \frac{Q}{I}\ \ {\rm and}\ \ P_U = \frac{U}{I} \;. 
\label{Eq_Pq_Pu_Def}
\end{equation}
$P_Q$ and $P_U$ were measured by combining the photon counts
(background subtracted) of ordinary and extra-ordinary beams
($f^\mathrm{o}$ and $f^\mathrm{e}$, respectively) observed at retarder
waveplate positions $\alpha = 0^\circ$, $22.5^\circ$, $45^\circ$, and
$67.5^\circ$, as given by the following formula:
\begin{equation}
\begin{array}{rcl}
P_Q & = & 
\frac{1}{2}
\Bigg\{
\left(\frac{f^\mathrm{o} - f^\mathrm{e}}{f^\mathrm{o} + 
            f^\mathrm{e}}\right)_{\alpha= 0^\circ} -
\left(\frac{f^\mathrm{o} - f^\mathrm{e}}{f^\mathrm{o} +
            f^\mathrm{e}}\right)_{\alpha=45^\circ}
\Bigg\} \\
P_U & = & 
\frac{1}{2}
\Bigg\{
\left(\frac{f^\mathrm{o} - f^\mathrm{e}}{f^\mathrm{o} +
f^\mathrm{e}}\right)_{\alpha=22.5^\circ} -
\left(\frac{f^\mathrm{o} - f^\mathrm{e}}{f^\mathrm{o} +
f^\mathrm{e}}\right)_{\alpha=67.5^\circ}
\Bigg\} \\
\end{array}
\label{Eq_Pq_Pu}
\end{equation}
(see FORS1/2 User manual, VLT-MAN-ESO-13100-1543).
The error on $P_Q$ or $P_U$ is 
\begin{equation}
\begin{array}{rcl}
\sigma^2_{P_X} & = &
  \left(\left(\frac{f^\mathrm{e}}{(f^\mathrm{o} + f^\mathrm{e})^2}\right)^2 \sigma^2_{f^\mathrm{o}} +
  \left(\frac{f^\mathrm{o}}{({f^\mathrm{o} + f^\mathrm{e}})^2}\right)^2
    \sigma^2_{f^\mathrm{e}}\right)_{\alpha=\phi_0 } + \\
             &   &
  \left(\left(\frac{f^\mathrm{e}}{({f^\mathrm{o} + f^\mathrm{e}})^2}\right)^2 \sigma^2_{f^\mathrm{o}} +
  \left(\frac{f^\mathrm{o}}{({f^\mathrm{o} + f^\mathrm{e}})^2}\right)^2
    \sigma^2_{f^\mathrm{e}}\right)_{\alpha=45^\circ + \phi_0} \;, \\\end{array}
\label{Eq_Sigma_QU}
\end{equation}
where $\phi_0 = 0^\circ$ if $X = Q$ and $\phi_0=22.5^\circ$
if $X = U$. If the polarization of the targets is always small, in order to give
an estimate of the quantities $\sigma_{P_X}$ one can substitute in
Eq.~(\ref{Eq_Sigma_QU})
$f^{\rm e} = f^{\rm o} = f$, with $f$ independent of $\alpha$. Also, if one can assume
$\sigma_{f^{\rm e}} = \sigma_{f^{\rm o}} = \sigma_{f}$ we have
\begin{equation}
\sigma_{P_X} = \frac{\sigma_{f}}{f} \;.
\end{equation}
If the error on the photon counts is entirely due to photon-counts, we get 
\begin{equation}
\sigma_{P_X} = \frac{1}{2}\,\frac{1}{\sqrt{f}} \;.
\end{equation}

From $P_Q$ and $P_U$ we have obtained the total
fraction of linear polarization $P_{\rm L}$ and the position angle $\theta$ (see
Landi Degl'Innocenti et al. 2007).  
For the cases where polarimetric errors are small with respect to the signal
($\sigma_{P_{\rm  Q}} \ll P_{\rm L}$, $\sigma_{P_{\rm  U}} \ll P_{\rm L}$),
the errors on $P_{\rm L}$ and $\theta$ are
\begin{equation}
\sigma_{P_{\rm L}} = \left[\cos^2(2 \theta) \sigma^2_{P_Q} +
                           \sin^2(2 \theta) \sigma^2_{P_U} \right]^{1/2}
\label{Eq_Err_P}
\end{equation}
\begin{equation}
\sigma_{\theta} = \frac{1}{2}
                  \frac{\left[\sin^2(2 \theta) \sigma^2_{P_Q} +
                              \cos^2(2 \theta) \sigma^2_{P_U}\right]^{1/2}}{P_{\rm L}} \;.
\label{Eq_Err_theta}
\end{equation}
Note that if $\sigma_{P_Q} =\sigma_{P_U}$ one gets
\begin{equation}
\sigma_{P_{\rm L}} = \sigma_{P_Q} =\sigma_{P_U} 
\end{equation}
and
\begin{equation}
\sigma_\theta = \frac{1}{2}\,\frac{\sigma_{P_L}}{P_{\rm L}} \;.
\end{equation}

\subsection{IPOL data}
All the science frames were bias subtracted using the corresponding 
master bias obtained from a series of five frames taken the morning
after the observations.  No flat-field correction was carried out as
it is irrelevant for the purpose of measuring the polarization through
Eq.~(\ref{Eq_Pq_Pu}).  The flux in the ordinary and extra-ordinary
beams was measured via simple aperture photometry, that was performed
using the {\tt DAOPHOT} package implemented in {\tt IRAF}.  Once the
fluxes for the ordinary and extraordinary beams for each retarder
waveplate position were measured, we used a dedicated C routine to
calculate $P'_Q$ and $P'_U$ through Eq.~(\ref{Eq_Pq_Pu}), the
corresponding errors via Eq.~(\ref{Eq_Sigma_QU}), then the fraction of
linear polarization $P'_{\rm L}$ and the position angle $\theta'$ with
Eqs.~(6) and (7) of Landi Degl'Innocenti et al. \cite{Lanetal07}, with
the corresponding errors given in Eqs.~(\ref{Eq_Err_P}) and
(\ref{Eq_Err_theta}) of this paper. To compensate a chromatism
problem of the half wave plate (see Sect.~3.2) a new position angle
$\theta$ was obtained as
\begin{equation}
\theta = \theta' - \epsilon_\theta({\rm F})
\label{Eq_Theta}
\end{equation}
where $\epsilon_\theta({\rm F})$ is a correction factor that depends
upon the filter F. The $\epsilon_\theta({\rm F})$ values
are tabulated in the FORS1/2 user manual
(see also\\ {\tt http://www.eso.org/instruments/fors/inst/pola.html}).
We obtained the final $P_Q$ and $P_U$ values as
\begin{equation}
\begin{array}{rcl}
P_Q &=& P'_{\rm L}\,\cos(2\theta) \\
P_U &=& P'_{\rm L}\,\sin(2\theta) \;. \\
\end{array}
\end{equation}
with $\theta$ given by Eq.~(\ref{Eq_Theta}) (note that $P_{\rm L} = P'_{\rm L}$).

\subsection{PMOS data}
PMOS data were pre-processed with the packages for spectra analysis
implemented in IRAF. Spectra in the ordinary and extraordinary beams
were bias-subtracted, then optimally extracted and wavelength
calibrated using IRAF routines, and finally processed using dedicated C
routines to calculate $P_Q$, $P_U$, $P_{\rm L}$, and $\theta$ with the
corresponding errors using Eqs.~(\ref{Eq_Pq_Pu}) and (\ref{Eq_Sigma_QU})
of this paper, and
Eqs.~(6) and (7) of Landi Degl'Innocenti et al. \cite{Lanetal07}.

\begin{figure}
\begin{center}
\plotfiddle{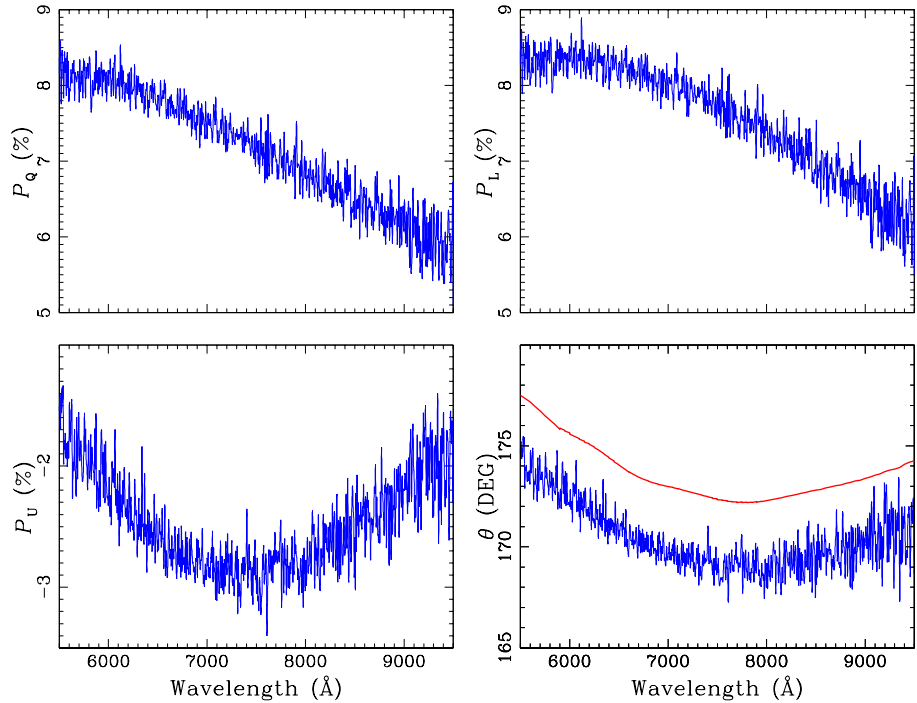}{8.5cm}{270}{50}{50}{-200}{270}
\end{center}
\caption{Star Ve 6-23: $P'_Q$, $P'_U$, $P_{\rm L}$, and $\theta'$ observed in the night
17 December 2003 with grism 150\,I and filter GG435. The non-constant $\theta$ value
is due to the chromatism of the half waveplate. The thin solid line shows the
offset angle $\epsilon_\theta(\lambda)$ (data are available at
{\tt http://www.eso.org/instruments/fors/inst/pola.html})
that has to be subtracted to the observed position angle to compensate for the 
waveplate chromatism (in the figure, $\epsilon_\theta(\lambda)$ is offset by
a constant to allows its visualization).}
\label{figure esempio}
\end{figure}

Figure \ref{figure esempio} shows the results for the star Ve 6-23
observed in the night 17 December 2003 with the 150\,I grism and the GG\,435
filter. The position angle of interstellar polarization is expected
to have a constant value independent of wavelength. The botton right
panel of Fig.~\ref{figure esempio} shows that this is not the case for
the star Ve 6-23 (as well as for all other stars of our sample). This is due
to a chromatism problem of the half-wave retarder waveplate already
discussed in Sect.~3.1.

To compare the IPOL with the PMOS observations we convolved the
polarized spectra obtained in PMOS with the transmission functions of
the \textit{BVRI} Bessel filters
used in IPOL mode.  For each filter F we calculated
\begin{equation}
P_{Q}({\rm F})=\frac{\int_{0}^{\infty}\mathrm{d}{\lambda}\,
P_Q({\lambda}) I_Q({\lambda}) T_{\rm F}({\lambda})}{\int_{0}^{\infty}\mathrm{d}{\lambda}\,I_Q({\lambda})T_{\rm F}({\lambda})}
\qquad 
P_{U}({\rm F})=\frac{\int_{0}^{\infty}\mathrm{d}{\lambda}\,
P_U({\lambda})I_U(\lambda) T_{\rm F}({\lambda})}{\int_{0}^{\infty}\mathrm{d}{\lambda}\,I_U({\lambda})T_{\rm F}({\lambda})}
\end{equation}
where $T_{\rm F}$ is the transmission function of the F filter, and
\begin{equation}
\begin{array}{rcl}
I_Q &=& \left(f^{\rm o} + f^{\rm e}\right) \vert_{\alpha = 0^\circ} + 
        \left(f^{\rm o} + f^{\rm e}\right) \vert_{\alpha = 45^\circ} \\[2mm]
I_U &=& \left(f^{\rm o} + f^{\rm e}\right) \vert_{\alpha = 22.5^\circ} + 
        \left(f^{\rm o} + f^{\rm e}\right) \vert_{\alpha = 67.5^\circ} \; .\\
\end{array}
\end{equation}
The polarization values so obtained
have been finally modified according to the procedure followed for the
IPOL data, in order to compensate for the chromatism of the half wave
plate.

\begin{figure}
\begin{center}
\plotfiddle{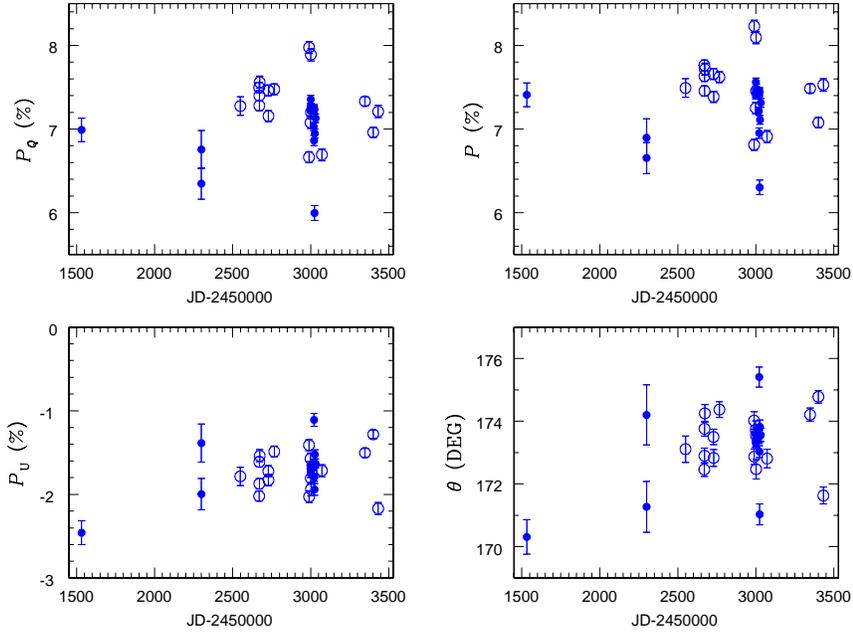}{7.5cm}{270}{45}{45}{-180}{250}
\end{center}
\caption{IPOL data (full circles) and rebinned PMOS data (open circles) for the
star Ve~6-23 in the \textit{B} band}
\label{figure misto Vela}
\end{figure}

\begin{figure}
\begin{center}
\plotfiddle{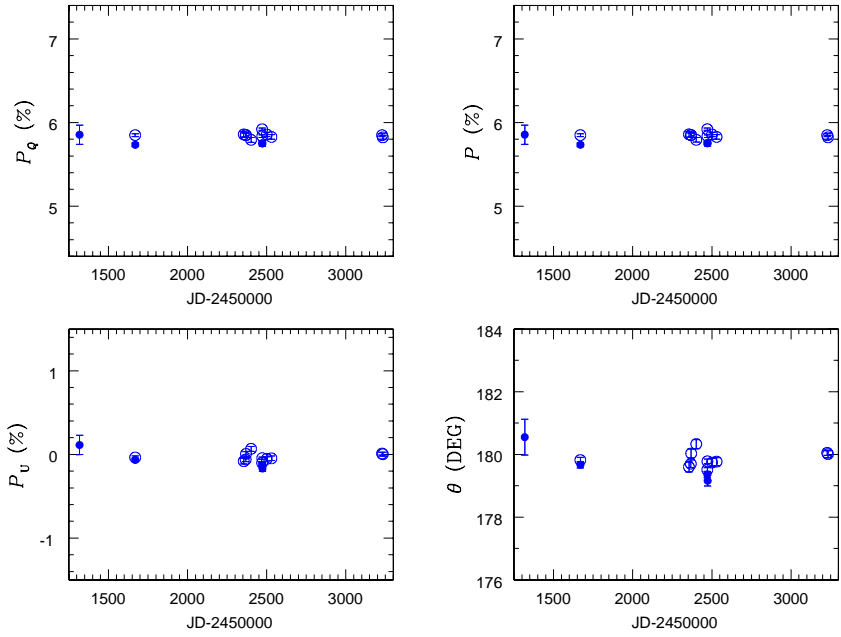}{7.5cm}{270}{45}{45}{-180}{250}
\end{center}
\caption{IPOL data (full circles) and rebinned PMOS data (open circles) for the
star CD\,$-$28~13479 in the \textit{B} band}
\label{figure misto CD}
\end{figure}

Error bars obtained in PMOS mode are smaller than those obtained in
IPOL mode. This is a consequence of the fact that the signal-to-noise
ratio of the observations is basically limited by the full well
capacity of the CCD pixels, or by the hardware limitations of the
digital-analogic converter. Rebinning spectropolarimetric data permits
one to integrate the signal over a much higher number of pixels than
possible in imaging mode, leading to a much higher signal-to-noise
ratio.  A comparison between IPOL and rebinned PMOS data can be seen
in Figs.~\ref{figure misto Vela} and \ref{figure misto CD} for stars
Ve~6-23 and CD\,$-$28~13479, respectively, in the \textit{B} band.

%
\section{Results and discussion}
Figures~\ref{figure misto Vela} and \ref{figure misto CD} show the
polarization observed in IPOL and in PMOS mode for stars Ve~6-23 and
CD~$-$28\,13479, respectively, plotted as a function of the
observation epoch. The plot scales are the same for both stars. The data for
the star Ve~6-23 appear much more scattered than those 
for the star CD~$-28$~13479. Whereas some scattering may well be due to
undetected instrument or data reduction problems, Fig.~\ref{figure misto
Vela} suggests that the polarization of the star Ve~6-23 may exhibit a
short-term variability. 

For each star, we have grouped all observations obtained with
similar instrument mode and with the same filter (or convolved with the
same trasmission function, in case of PMOS observations). From each
group so obtained  we have
calculated the medians of the observed $P_Q$
and $P_U$ values, $\widetilde{P_X}$ (with $X=Q$ and $U$),
and the median absolute deviations (MAD), i.e., the medians of
the distributions
\begin{equation}
\vert (P_X)_{i} - \widetilde{P_X}  \vert \;.
\end{equation}
Setting $\sigma = 1.48$\,MAD (e.g., Huber 1981,
pp.~107--108), we have then rejected those $(P_X)_{i}$ values for which
\begin{equation}
\vert (P_X)_{i} - \widetilde{P_X} \vert > 3\,\sigma
\end{equation}
Finally, from the remaining values, we have calculated the weighted
averages $\widehat{P_Q}$ and $\widehat{P_U}$:
\begin{equation}
\widehat{P_X} = \frac{\sum_i \frac{(P_X)_i}{(\sigma^2_X)_i}}{\sum_i \frac{1}{(\sigma^2_X)_i}} \;.  
\end{equation}
 To each average value,
we have associated the error given by
\begin{equation}
\sigma^2_{P_X} =
\frac{1}{N-1}\,
\frac{\sum_i\frac{\left( (P_X)_i - \widehat{P_X} \right)^2}{(\sigma^2_{P_X})_i}}{{ \sum_i \frac{1}{(\sigma^2_{P_X})_i} }}
\label{Eq_Erroruccio}
\end{equation}

The results are shown in Table~\ref{table after sigma clipping
polarized} (for the stars with high polarization signal) and
\ref{table after sigma clipping not polarized} (for the stars with low
polarization signal). Stars observed only once were not included in
the tables. For stars with less than four observations we did not run
the $k\sigma$-clipping algorithm.

\subsection{Polarized stars}
Table~\ref{table after sigma clipping polarized} may be used as a
reference to check instrument performance and stability within the
quoted error bars, both in IPOL and PMOS mode. Note that relatively
large errors may point to a star's instrisic variability, especially
if large errors are associated to large data sets. Observations within
similar filter bands are fairly, but not always fully consistent, see
Sect.~4.2 below.

It should be noted that the actual correction that has to be applied
to the broadband polarization measurements to compensate for the
chromatism of the retarder waveplate depends on the shape of the
star's spectral energy distribution (convolved with the transmission
of the telescope optics). In fact, we have applied a correction that
is independent of the star's colour. This is probably the reason why
the position angles of the stars of Table~\ref{table after sigma
clipping polarized} are slightly filter dependent. For the same
reason, caution should be adopted when comparing the results reported
in Table~\ref{table after sigma clipping polarized} with polarimetric
observations obtained with other instruments.

\subsection{Unpolarized stars}
All stars observed in IPOL mode and reported in Table~\ref{table after
sigma clipping not polarized} have $P_Q$ and $P_U$ values consistent
with zero. This means that all these stars may be considered as
unpolarized standard stars within a typical accuracy better than
$3 \times 10^{-4}$. At the same time, the available IPOL observations do not
show evidence for significant instrumental polarization \textit{in the
center of the instrument field of view}\footnote{For a study of the
FORS1 instrumental polarization off-axis see Patat \& Romaniello
\cite{PatRom06}}.

WD~2007-303, that was observed in PMOS mode only, is polarized at
about 0.5\,\% level in $P_Q$, and should not be used as unpolarized
standard star.

The remaining PMOS observations hint that there is a small
instrumental offset in $P_Q$. The weigthed average of all $P_Q$ values
of Table~\ref{table after sigma clipping not polarized} (but without
considering the star WD~2007-303) is $P_Q(B)\vert_0 = 0.07 \pm 0.01$\,\%,
and $P_Q(V)\vert_0 = 0.09 \pm 0.01$\,\% for the $B$ and the $V$ filter,
respectively (the errors were calculated with
Eq.~(\ref{Eq_Erroruccio})). The averages of all $P_U$
values of Table~\ref{table after sigma clipping not polarized} is
fully consistent with 0 (with an error bar of about 0.005\,\%) both in
the $B$ and in the $V$ filters. The observed $P_Q$ offset, that
appears instrinsic to the PMOS mode only, may be associated to some
but not all grism+filter combinations, and deserves further
investigation.

\acknowledgements L.~Fossati acknowledges ESO DGDF for a four month 
studentship at ESO Santiago/Vitacura.

\newpage
\begin{scriptsize}
\begin{longtable}{cccr@{$\pm$}rr@{$\pm$}rr@{$\pm$}rr@{$\pm$}rcc}
\caption[ ]{\label{table after sigma clipping polarized} 
Weighted averages of the observations of polarized stars in IPOL and
PMOS mode with various instrument settings. The averages were
calculated after running the $k\sigma$-clipping procedure
explained in the text.  $N_{\rm s}$ and $N_{\rm e}$ represent the number of data
before and after running the $k\sigma$-clipping
procedure, respectively. Note that for $N \le 3$ no
$k\sigma$-clipping procedure was applied.}\\
\hline\hline
Star  & Mode  & Band  & 
\multicolumn{2}{c}{$P_Q$ (\%)} & 
\multicolumn{2}{c}{$P_U$ (\%)} &
\multicolumn{2}{c}{$P$   (\%)} &
\multicolumn{2}{c}{$\theta$}   & 
$N_{\rm s}$ & $N_{\rm e}$     \\
\hline
\endfirsthead
\caption{continued.}\\
\hline\hline
Star  & Mode  & Band  & 
\multicolumn{2}{c}{$P_Q$ (\%)} & 
\multicolumn{2}{c}{$P_U$ (\%)} &
\multicolumn{2}{c}{$P$   (\%)} &
\multicolumn{2}{c}{$\theta$}   & 
$N_{\rm s}$ & $N_{\rm e}$     \\
\hline
\endhead
\hline
\endfoot
HD 245310   &{\tt PMOS}&$V$&$-$1.02 & 0.02 &$-$3.98 & 0.01 & 4.10 & 0.01 & 127.79 & 0.12 & 2  & 2  \\[2mm]
NGC 2024 1  &{\tt IPOL}&$B$&   0.80 & 0.07 &$-$8.28 & 0.01 & 8.31 & 0.01 & 137.77 & 0.24 & 6  & 4  \\
	    &{\tt IPOL}&$V$&   0.16 & 0.20 &$-$9.64 & 0.06 & 9.65 & 0.06 & 135.47 & 0.59 & 3  & 3  \\
	    &{\tt IPOL}&$R$&   0.28 & 0.01 &$-$9.62 & 0.01 & 9.62 & 0.01 & 135.84 & 0.02 & 4  & 3  \\
	    &{\tt IPOL}&$I$&   0.40 & 0.05 &$-$9.11 & 0.04 & 9.12 & 0.04 & 136.26 & 0.15 & 2  & 2  \\[1mm]
	    &{\tt PMOS}&$B$&   1.05 & 0.09 &$-$8.02 & 0.06 & 8.09 & 0.06 & 138.75 & 0.32 & 10 & 9  \\
	    &{\tt PMOS}&$V$&   0.58 & 0.05 &$-$9.51 & 0.02 & 9.53 & 0.02 & 136.74 & 0.16 & 10 & 9  \\[2mm]
Ve 6-23     &{\tt IPOL}&$B$&   7.12 & 0.05 &$-$1.66 & 0.02 & 7.32 & 0.05 & 173.42 & 0.10 & 14 & 11 \\
	    &{\tt IPOL}&$V$&   7.91 & 0.05 &$-$2.38 & 0.06 & 8.26 & 0.05 & 171.61 & 0.21 & 12 & 10 \\
	    &{\tt IPOL}&$R$&   7.56 & 0.06 &$-$2.32 & 0.03 & 7.90 & 0.05 & 171.47 & 0.13 & 13 & 12 \\
	    &{\tt IPOL}&$I$&   6.88 & 0.06 &$-$2.28 & 0.08 & 7.25 & 0.07 & 170.86 & 0.31 & 5  & 4  \\[1mm]
	    &{\tt PMOS}&$B$&   7.24 & 0.08 &$-$1.68 & 0.06 & 7.43 & 0.08 & 173.48 & 0.23 & 17 & 17 \\
	    &{\tt PMOS}&$V$&   7.98 & 0.02 &$-$2.29 & 0.04 & 8.30 & 0.02 & 171.99 & 0.12 & 13 & 13 \\
	    &{\tt PMOS}&$R$&   7.76 & 0.01 &$-$2.24 & 0.01 & 8.08 & 0.01 & 171.97 & 0.03 & 2  & 2  \\
	    &{\tt PMOS}&$I$&   7.02 & 0.06 &$-$2.01 & 0.02 & 7.31 & 0.06 & 172.02 & 0.11 & 6  & 5  \\[2mm]
HD 111579   &{\tt PMOS}&$B$&$-$5.21 & 0.18 &$-$2.23 & 0.56 & 5.67 & 0.29 & 101.60 & 2.56 & 2  & 2  \\
 	    &{\tt PMOS}&$V$&$-$5.73 & 0.03 &$-$2.49 & 0.21 & 6.25 & 0.10 & 101.76 & 0.87 & 2  & 2  \\[2mm]
CD-28 13479 &{\tt IPOL}&$B$&   5.70 & 0.01 &$-$0.11 & 0.03 & 5.70 & 0.01 & 179.47 & 0.13 & 4  & 3  \\
      	    &{\tt IPOL}&$V$&   6.24 & 0.03 &$-$0.18 & 0.04 & 6.25 & 0.03 & 179.18 & 0.20 & 6  & 6  \\
      	    &{\tt IPOL}&$R$&   6.07 & 0.02 &$-$0.13 & 0.02 & 6.07 & 0.02 & 179.39 & 0.10 & 15 & 11 \\[1mm]
      	    &{\tt PMOS}&$B$&   5.80 & 0.01 &$-$0.03 & 0.01 & 5.80 & 0.01 & 179.85 & 0.05 & 11 & 9  \\
      	    &{\tt PMOS}&$V$&   6.31 & 0.01 &$-$0.16 & 0.02 & 6.32 & 0.01 & 179.27 & 0.07 & 14 & 13 \\
      	    &{\tt PMOS}&$I$&   5.61 & 0.04 &$-$0.16 & 0.02 & 5.61 & 0.04 & 179.18 & 0.11 & 2  & 2  \\[2mm]
HD 316232   &{\tt IPOL}&$V$&   4.85 & 0.02 &   0.41 & 0.07 & 4.87 & 0.02 &   2.40 & 0.38 & 3  & 3  \\
       	    &{\tt IPOL}&$R$&   4.69 & 0.03 &   0.55 & 0.02 & 4.72 & 0.03 &   3.36 & 0.12 & 3  & 3  \\
       	    &{\tt IPOL}&$I$&   4.32 & 0.01 &   0.40 & 0.04 & 4.34 & 0.01 &   2.65 & 0.26 & 2  & 2  \\[1mm]
       	    &{\tt PMOS}&$B$&   4.62 & 0.02 &   0.57 & 0.01 & 4.66 & 0.02 &   3.54 & 0.06 & 4  & 4  \\
       	    &{\tt PMOS}&$V$&   4.97 & 0.01 &   0.51 & 0.02 & 5.00 & 0.01 &   2.94 & 0.10 & 3  & 3  \\[2mm]
BD-14 4922  &{\tt IPOL}&$B$&$-$0.98 & 0.03 &   5.63 & 0.01 & 5.71 & 0.01 &  49.93 & 0.15 & 4  & 4  \\
 	    &{\tt IPOL}&$V$&$-$0.99 & 0.02 &   6.07 & 0.02 & 6.15 & 0.02 &  49.62 & 0.08 & 5  & 5  \\
 	    &{\tt IPOL}&$R$&$-$0.98 & 0.05 &   5.69 & 0.07 & 5.77 & 0.07 &  49.90 & 0.24 & 2  & 2  \\[1mm]
 	    &{\tt PMOS}&$B$&$-$0.97 & 0.01 &   5.66 & 0.01 & 5.74 & 0.01 &  49.88 & 0.07 & 5  & 4  \\
 	    &{\tt PMOS}&$V$&$-$0.93 & 0.01 &   6.06 & 0.01 & 6.13 & 0.01 &  49.35 & 0.03 & 5  & 3  \\
 	    &{\tt PMOS}&$I$&$-$0.70 & 0.03 &   5.07 & 0.01 & 5.12 & 0.01 &  48.92 & 0.19 & 2  & 2  \\[2mm]
BD-12 5133  &{\tt IPOL}&$B$&   1.87 & 0.04 &$-$3.95 & 0.05 & 4.37 & 0.05 & 147.64 & 0.28 & 6  & 5  \\
 	    &{\tt IPOL}&$V$&   1.75 & 0.04 &$-$4.00 & 0.04 & 4.37 & 0.04 & 146.84 & 0.25 & 5  & 4  \\
 	    &{\tt IPOL}&$R$&   1.63 & 0.02 &$-$3.68 & 0.02 & 4.02 & 0.02 & 146.97 & 0.13 & 16 & 13 \\
 	    &{\tt IPOL}&$I$&   1.10 & 0.29 &$-$3.39 & 0.07 & 3.57 & 0.09 & 143.99 & 2.27 & 3  & 3  \\[1mm]
 	    &{\tt PMOS}&$B$&   1.94 & 0.04 &$-$3.94 & 0.03 & 4.39 & 0.03 & 148.09 & 0.23 & 5  & 4  \\
 	    &{\tt PMOS}&$V$&   1.80 & 0.04 &$-$4.05 & 0.01 & 4.43 & 0.02 & 146.99 & 0.22 & 6  & 4  \\
 	    &{\tt PMOS}&$I$&   1.03 & 0.47 &$-$3.37 & 0.06 & 3.53 & 0.11 & 143.45 & 3.73 & 2  & 2  \\[2mm]
BD-13 5073  &{\tt IPOL}&$R$&   2.10 & 0.01 &$-$2.99 & 0.03 & 3.66 & 0.02 & 152.55 & 0.11 & 2  & 2  \\[1mm]
 	    &{\tt PMOS}&$V$&   3.40 & 1.59 &$-$2.64 & 1.17 & 4.30 & 1.47 & 161.11 & 8.77 & 2  & 2  \\
\hline
\end{longtable}

\end{scriptsize}
\newpage

\begin{scriptsize}
\begin{longtable}{cccr@{$\pm$}rr@{$\pm$}rcc}
\caption[ ]{\label{table after sigma clipping not polarized} 
Weighted averages of the observations of unpolarized stars in
IPOL and PMOS mode with various instrument settings. The meaning
of the various columns is the same as in 
Table~\ref{table after sigma clipping polarized}}\\
\hline\hline
Star  & Mode  & Band  & 
\multicolumn{2}{c}{$P_Q$ (\%)} &
\multicolumn{2}{c}{$P_U$ (\%)} &
$N_{\rm s}$ & $N_{\rm e}$     \\
\hline
\endfirsthead
\caption{continued.}\\
\hline\hline
Star  & Mode  & Band  & 
\multicolumn{2}{c}{$P_Q$ (\%)} &
\multicolumn{2}{c}{$P_U$ (\%)} &
$N_{\rm s}$ & $N_{\rm e}$     \\
\hline
\endhead
\hline
\endfoot
WD 2359-434&{\tt IPOL}&$R$&   0.03 & 0.06 &$-$0.20 & 0.09 & 4  & 4  \\
 	   &{\tt IPOL}&$I$&$-$0.00 & 0.01 &   0.03 & 0.04 & 4  & 4  \\[2mm]
HD 10038   &{\tt PMOS}&$B$&   0.07 & 0.01 &$-$0.03 & 0.02 & 3  & 3  \\
 	   &{\tt PMOS}&$V$&   0.11 & 0.01 &$-$0.03 & 0.01 & 2  & 2  \\[2mm]
HD 13588   &{\tt PMOS}&$B$&   0.10 & 0.01 &   0.00 & 0.01 & 4  & 4  \\
 	   &{\tt PMOS}&$V$&   0.10 & 0.01 &   0.00 & 0.01 & 4  & 4  \\[2mm]
WD 0310-688&{\tt IPOL}&$V$&   0.01 & 0.04 &$-$0.05 & 0.10 & 3  & 3  \\
 	   &{\tt IPOL}&$R$&   0.04 & 0.09 &   0.01 & 0.14 & 2  & 2  \\[2mm]
HD 42078   &{\tt PMOS}&$B$&   0.04 & 0.01 &$-$0.01 & 0.01 & 4  & 2  \\
 	   &{\tt PMOS}&$V$&   0.07 & 0.01 &$-$0.02 & 0.01 & 6  & 6  \\[2mm]
HD 64299   &{\tt PMOS}&$B$&$-$0.02 & 0.12 &$-$0.04 & 0.04 & 2  & 2  \\
 	   &{\tt PMOS}&$V$&$-$0.05 & 0.13 &$-$0.03 & 0.04 & 2  & 2  \\[2mm]
WD 0752-676&{\tt IPOL}&$B$&$-$0.00 & 0.01 &   0.06 & 0.03 & 4  & 3  \\
 	   &{\tt IPOL}&$V$&   0.00 & 0.01 &   0.03 & 0.05 & 3  & 3  \\[2mm]
HD 97689   &{\tt PMOS}&$B$&   0.16 & 0.04 &   0.01 & 0.01 & 3  & 3  \\
 	   &{\tt PMOS}&$V$&   0.14 & 0.08 &$-$0.00 & 0.01 & 2  & 2  \\[2mm]
WD 1615-154&{\tt IPOL}&$V$&   0.03 & 0.24 &$-$0.05 & 0.03 & 2  & 2  \\
       	   &{\tt IPOL}&$R$&$-$0.01 & 0.04 &$-$0.02 & 0.12 & 2  & 2  \\[1mm]
       	   &{\tt PMOS}&$B$&   0.07 & 0.02 &$-$0.03 & 0.02 & 5  & 5  \\
       	   &{\tt PMOS}&$V$&   0.13 & 0.01 &$-$0.02 & 0.02 & 4  & 4  \\[2mm]
WD 1620-391&{\tt IPOL}&$B$&   0.02 & 0.03 &$-$0.02 & 0.02 & 3  & 3  \\
       	   &{\tt IPOL}&$V$&$-$0.01 & 0.01 &$-$0.01 & 0.01 & 4  & 3  \\
       	   &{\tt IPOL}&$R$&$-$0.00 & 0.01 &   0.01 & 0.01 & 5  & 4  \\
       	   &{\tt IPOL}&$I$&   0.01 & 0.01 &   0.01 & 0.01 & 4  & 2  \\[1mm]
       	   &{\tt PMOS}&$B$&   0.06 & 0.02 &   0.00 & 0.01 & 2  & 2  \\
       	   &{\tt PMOS}&$V$&   0.08 & 0.02 &   0.00 & 0.01 & 2  & 2  \\[2mm]
HD 176425  &{\tt PMOS}&$V$&   0.15 & 0.13 &$-$0.08 & 0.06 & 2  & 2  \\[2mm]
WD 2007-303&{\tt PMOS}&$B$&   0.44 & 0.01 &$-$0.09 & 0.01 & 4  & 3  \\
 	   &{\tt PMOS}&$V$&   0.51 & 0.02 &$-$0.05 & 0.01 & 4  & 4  \\[2mm]
WD 2039-202&{\tt IPOL}&$R$&   0.19 & 0.23 &$-$0.14 & 0.40 & 2  & 2  \\[1mm]
       	   &{\tt PMOS}&$B$&   0.03 & 0.03 &   0.00 & 0.01 & 2  & 2  \\
       	   &{\tt PMOS}&$V$&   0.03 & 0.01 &$-$0.00 & 0.01 & 2  & 2  \\
       	   &{\tt PMOS}&$I$&   0.10 & 0.01 &   0.01 & 0.01 & 2  & 2  \\[2mm]
WD 2149+021&{\tt IPOL}&$V$&   0.01 & 0.03 &$-$0.04 & 0.01 & 2  & 2  \\
       	   &{\tt IPOL}&$R$&   0.14 & 0.14 &$-$0.02 & 0.09 & 3  & 3  \\[1mm]
       	   &{\tt PMOS}&$V$&   0.09 & 0.01 &$-$0.01 & 0.03 & 2  & 2  \\
       	   &{\tt PMOS}&$I$&   0.16 & 0.01 &$-$0.01 & 0.04 & 2  & 2  \\
\hline
\end{longtable}

\end{scriptsize}

\end{document}